\documentstyle[12pt,epsfig,cite]{article}
\newcommand{\beq}{\begin{equation}}
\newcommand{\eeq}{\end{equation}}
\newcommand{\bea}{\begin{eqnarray}}
\newcommand{\eea}{\end{eqnarray}}

\newcommand{\gsim}{\lower.7ex\hbox{$
\;\stackrel{\textstyle>}{\sim}\;$}}
\newcommand{\lsim}{\lower.7ex\hbox{$
\;\stackrel{\textstyle<}{\sim}\;$}}

\def\op{{\bf P}}

\def\cp{{\bf CP}}

\begin{document}
\thispagestyle{empty}
\vspace*{-22mm}

\begin{flushright}
UND-HEP-06-BIG\hspace*{.08em}04\\
hep-ph/0603234\\

\end{flushright}
\vspace*{1.3mm}

\begin{center}
{\Large{\bf
A SEND-OFF AFTER DIF06: WHAT DO WE \\
\vspace{3mm}
NEED TO KNOW TO UNDERSTAND MORE?
}}
\vspace*{19mm}

{\Large{\bf I.I.~Bigi}} \\
\vspace{7mm}

{\sl Department of Physics, University of Notre Dame du Lac}
\vspace*{-.8mm}\\
{\sl Notre Dame, IN 46556, USA}\\
{\sl email: ibigi@nd.edu}
\vspace*{10mm}

{\bf Abstract}\vspace*{-1.5mm}\\
\end{center}

\noindent 
After a brief look back at the roles played by hadronic machines and $e^+e^-$ colliders I 
emphasize that continuing dedicated studies of heavy flavour transitions should be central 
to our efforts of decoding nature's `Grand Design'. For studies `instrumentalizing' the high 
sensitivity of \cp~violation will presumably be 
essential to identify salient features of the New Physics anticipated for the TeV scale 
and hopefully discovered directly at the LHC. An $e^+e^-$ Super-Flavour 
Factory would provide the optimal platform for such a program. 

\vspace{2cm}

This is {\em not} a summary -- to my considerable relief I was asked not to give one. These are not 
{\em the} conclusions either. You can never give the conclusions when the `boss', in this case M. Calvetti, is speaking right after you. 
\footnote{A Justice on the US Supreme Court once said: "We are not the 
Supreme Court, because we are infallible. We are infallible, since we are the Supreme (i.e. final) Court." Working at a Catholic University I am not unfamiliar with this issue.} Instead I will offer merely my 
personal reflections. 

\section{A Short Look Back}
\label{BACK}
Let us look at two exhibits from the past. 

{\em Exhibit A:}

\noindent The weak bosons were first found at the CERN SPPS -- a hadronic collider. It was  
LEP I \& II -- an $e^+e^-$ collider -- that established the electroweak gauge sector of the 
Standard Model (SM) with a {\em quantitative} accuracy {\em well beyond} 
original expectations. 

{\em Exhibit B:} 

\noindent Indirect and direct \cp~violation were first uncovered at BNL, CERN and FNAL -- all 
hadronic machines. The $B$ factories at KEK and SLAC -- $e^+e^-$ machines -- 
established the Yukawa sector of the SM: CKM dynamics are able to 
describe (at least almost) all \cp~violation as observed in particle decays to a 
{\em quantitative} accuracy {\em well beyond} original expectations. 
The CKM paradigm has thus become a 
{\em tested} theory; \cp~violation has been `demystified' in the sense that if the dynamics are sufficiently 
rich to be able to support \cp~violation the latter can be large; this demystification will be completed 
once \cp~violation is found anywhere in the lepton sector. 

Let us remember also what happened with strangeness changing transitions: 
(i) The $\tau-\theta$ puzzle lead to the realization that \op~symmetry does not hold in the weak interactions. (ii) The observation that the production rate exceeds the decay rate by 
several orders of magnitude (giving rise to the name `strangeness') lead to the 
notion of associated production and in due course to the concept of quark families. 
(iii) The huge suppression of strangeness changing neutral currents -- as inferred from the tiny size 
of $\Delta M_K$ and of BR$(K_L \to \mu^+\mu^-)$ -- encouraged 
some daring minds to speculate about charm quarks. (iv) The observation of 
$K_L \to \pi^+\pi^-$ revealed \cp~violation and suggested the existence of a third quark 
family. 

All these features, which are now pillars of the SM, were New Physics {\em at that time}! 

\section{What to Do?}

\begin{figure}[t]
\vspace{7.0cm}
\includegraphics{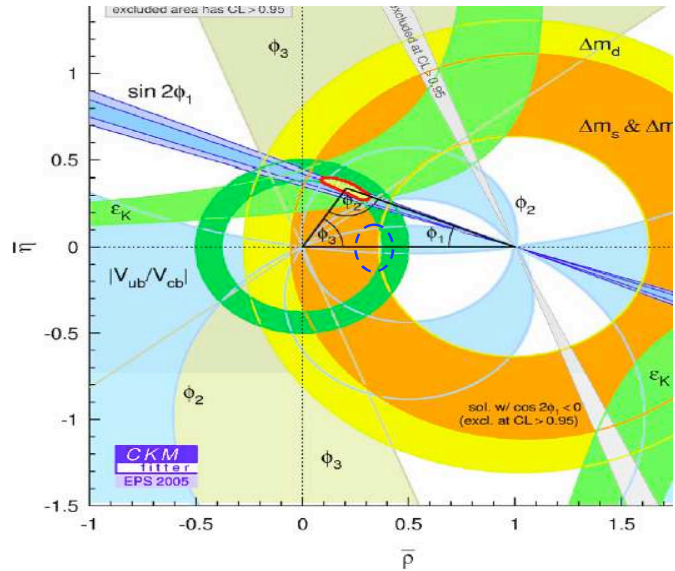}
 \caption{\it
      Constraints on the CKM triangle as of the end of 2005. 
    \label{CKM05} }
\end{figure}

The SM is a very peculiar creation. On one hand more and more victories have been attached to its 
banner. The years around the turn of the millenium have witnessed novel successes in its description 
of heavy flavour dynamics: most spectacularly the `Paradigm of truly large \cp~violation in $B$ decays' 
has been verified. We can even see the next triumph of CKM theory emerging: Consider the constraints 
on the CKM triangle as of the end of 2005 shown in Fig.\ref{CKM05}. 
If you remove the constraints from \cp~violation, in particular from $|\epsilon_K|$ and sin$2\phi_1$ 
(as well as from $\phi_{2,3}$), thus retaining only those from $|V(ub)/V(cb)|$ and $\Delta M(B)$, then a 
completely flat triangle -- i.e. one with zero area -- while not favoured, is not ruled out, as indicated 
by the broken circle. 

\begin{figure}[t]
\vspace{7.0cm}
\includegraphics{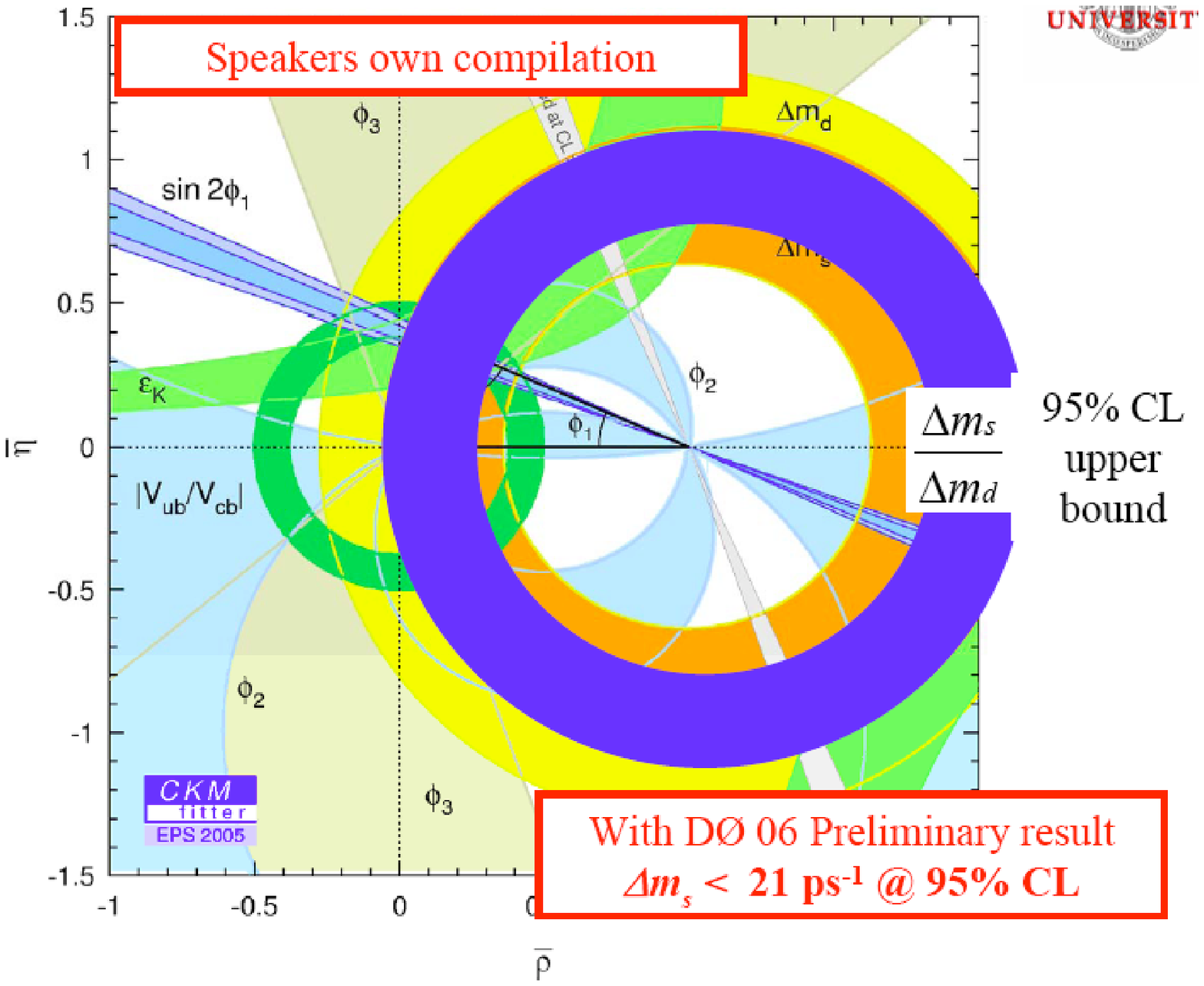}
 \caption{\it
      Constraints on the CKM triangle as of the end of 2005. 
    \label{CKM06} }
\end{figure}

{\em If} -- and that is admittedly a nontrivial `if' -- the 
evidence presented this spring by D0 \cite{D0} for an {\em upper} 
bound on the rate of $B_s$ oscillations  holds up -- $\Delta M(B_s) \leq 21$ p$^{-1}$  -- 
one has a qualitatively new scenario as can be seen from Fig.\ref{CKM06} \cite{CASEY}: 
\begin{itemize}
\item
\cp~{\em in}sensitive observables -- namely $|V(ub)/V(cb)|$ and $\Delta M(B_{d,s})$ -- imply 
a CKM triangle of {\em non-}zero area, i.e. the existence of \cp~violation. 
\item 
This triangle is fully consistent with the observed \cp~violation as expressed through 
$\epsilon_K$ and sin$2\phi_1$. 
\end{itemize}

Yet {\em none} of the successes of the SM -- including those just sketched -- 
invalidate the arguments for it 
being incomplete. This represents a virtual consensus in the community. 
New Physics is expected at the TeV scale. 

A clear majority opinion holds that we need more answers from nature to figure 
out the variant(s) of New Physics realized in nature and to establish that we are 
`on the right track'. After all `experiment is the Supreme Court of Physics'. 

Given the past history of searches for Physics beyond the SM this might remind you of 
Samuel Beckett's dictum: 
\begin{center}
Ever tried? Ever failed? \\
No matter. \\
Try again. Fail again. Fail better.
\end{center} 
But we can cheer up -- we know there is New Physics: With CKM dynamics being utterly irrelevant for 
baryogenesis, there has to be a `New \cp~Paradigm'. Likewise the SM cannot account for 
neutrino oscillations, dark matter and dark energy. In addition there are the SM's well known 
explanatory deficits. Thus we will not fail forever. 

Yet history rarely if ever repeats itself in an identical manner. Recognizing that the SM 
has succeeded in putting a vast array of phenomena occurring at very diverse scales under its 
roof, we cannot {\em count} on {\em numerically} massive manifestations of New Physics in heavy flavour transitions -- unlike what happened in the physics of strangeness before the era of the SM. 
It appears that nature has read the SM's book on flavour changing neutral currents 
-- at least for {\em down}-type quarks. 

Accordingly we need some luck to find this New Physics. Being lucky has of course to be 
part of the job description for a high energy physicist (in particular of the 
experimental variety).  Accuracy in acquiring and interpreting data will do wonders for 
enhancing our luck. 

\section{The Future}

I do not intend to beat around the bushes: in my view studies 
of heavy flavour dynamics 
\begin{itemize}
\item 
are of fundamental importance, 
\item 
their lessons cannot be obtained any other way and 
\item 
they can{\em not} become obsolete. 

\end{itemize}
I.e., no matter what studies of high $p_{\perp}$ physics at FNAL and the LHC will or 
will not show -- and I am confident they will show a lot -- detailed and comprehensive 
analyses of flavour physics will remain crucial in our efforts to reveal 
`Nature's Grand Design'. 

I see three possible scenarios of the next five to eight years: 

Scenario A -- the `optimal' one: 

\noindent New Physics has been observed at high $p_{\perp}$. It is then 
{\em mandatory} to study 
their impact on flavour dynamics, which is greatly facilitated by the mass scales  
of the New Physics being known. Even a negative  result -- i.e. no discernible 
impact on heavy flavour decays -- would be a highly important result in this scenario, 
however frustrating it would be for our experimental colleagues. 

Scenario B -- the `intriguing' one: 

\noindent Deviations from the SM have been established in heavy flavour decays. 
Recently considerable excitement has been created by the `$b \to s \bar ss$ anomaly', i.e. 
by experimental evidence that modes driven by this effective `Penguin-like' operator exhibit 
markedly lower \cp~asymmetries than predicted by the SM. While the spectacular discrepancies have  faded away, some have remained \cite{BABAR}, and I find those actually more believable than the original ones. 
Therefore  we better keep a close watch on them, 

Scenario C -- the `frustrating' one: 

\noindent {\em No further} deviation from the SM has been established at high or low 
$p_{\perp}$. 

While I am optimistic it will turn out to be Scenario A, I would like to emphasize that none of these 
scenarios weakens the importance of continuing heavy flavour studies for our quest of 
finding out about nature's basic forces. 

Yet we do not live and work in isolation. Following what was said by Nakada at this meeting 
I would like 
to formulate a `Generalized Nakada Concern': While more than ever before 
we have many promising avenues for exploring fundamental physics, while we have more 
technical tools and capabilities than ever, we live in a world with immense political, 
social and environmental problems; furthermore we have to deal with governments with less interest 
in basic research to a degree that goes well beyond a justified pre-occupation 
with these problems. How do we choose our priorities? 

This is an excellent question, and I do not have a general answer to it. I can offer 
only one criterion, namely to aim for {\em comprehensive} research goals. In the case under 
discussion here I want to emphasize that a Super-B factory is also a Super-Tau as well as 
Super-Charm factory of truly unique capabilities. It allows precise studies of a {\em third} family 
{\em down-}type {\em quark}, a  {\em third} family {\em down-}type {\em lepton} and a
{\em second} family {\em up-}type {\em quark}. In this context the studies of \cp~violation, 
oscillations and rare decays are {\em instrumentalized} to probe TeV scale New Physics. 

To achieve 
\begin{itemize} 
\item 
a luminosity of $10^{36}$ cm$^{-2}$ s$^{-1}$ 
\item 
with tiny beams and a hermetic detector 
allowing to study transitions with large amount of `invisible' energy -- like 
$B\to \tau ^+\tau ^-$, $\tau \nu$, $\tau \nu X$, $\tau \to l \nu \bar \nu$ etc. -- 
\item 
maybe even 
with one beam being polarized, 
\item 
`soon' and 
\item 
`here', i.e. near Rome -- 
\end{itemize}
to me sounds better than paradise. There is the promise that such a 
Super-Flavour Factory can be run not only at the $\Upsilon (4S)$ and 
$\Upsilon (5S)$ resonances, but even at much lower energies close to charm and $\tau$  
thresholds at an affordable cost in luminosity. While the primary goals in $\tau$ and charm 
physics -- searching for and probing \cp~violation as well as rare decays -- can profitably be 
pursued at the $\Upsilon (4S)$, future studies might show that lower energies are optimal 
for dealing with certain backgrounds. The statistically as well as systematically ambitious goals 
that need to be pursued in $B$ physics have been listed repeatedly at this workshop. Let me 
remind you also of equally challenging goals in $\tau$ and charm  studies 
\cite{CHARM}, namely to go after 
\cp~asymmetries as small as ${\cal O}(10^{-3})$ or even smaller.

It has often been noted that when `all is said and done' usually a lot more is said than done. 
We are seeing how the dream of a Super-Flavour Factory is turning into a vision. A whole lot 
needs to be done before it can be transformed into a project and finally elevated into reality.  

\begin{figure}[t]
\vspace{4.5cm}
\includegraphics{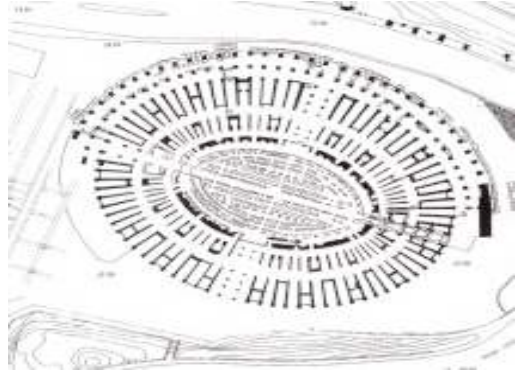}
 \caption{\it
      A prominent machine in Rome.
    \label{COLLISEO} }
\end{figure}

\begin{figure}[t]
\vspace{5.0cm}
\includegraphics{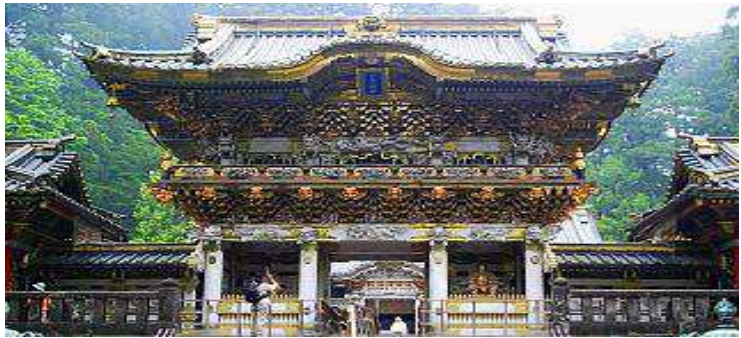}
\caption{\it The Baroque splendour of Nikko.
    \label{KEK} }
\end{figure}

I had mentioned in my opening remarks at this workshop \cite{OPEN} 
that the area around Rome has a long tradition of 
{\em linear} structures and shown you a historical example. In Biagini's talk 
\cite{BIAGINI} we have seen a 
brand new and very different  
`Solution 2' for an ILC-inspired Super-Flavour Factory. The fascination of Rome is such that you 
can find an example for almost anything in its rich heritage. `Solution 2' reminds me of the almost 2000 
years old structure shown in Fig. \ref{COLLISEO} full of tunnels with round as well as straight beam lines. 
Another intriguing aspect of it is that such a design would be portable to other places -- like 
KEK, which has its own heritage, see Fig.\ref{KEK}. 

\section{A Final Thought}

\begin{figure}[t]
\vspace{4.0cm}
\includegraphics{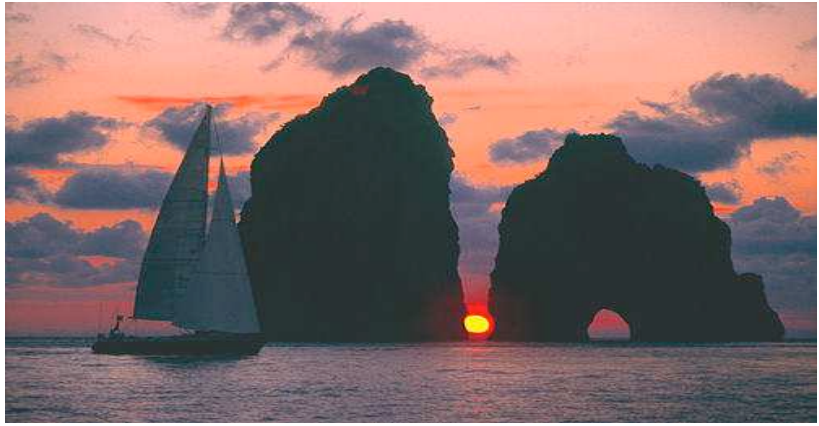}
 \caption{\it
      Another allegory on HEP's future landscape. 
    \label{FUTURE2} }
\end{figure}

In my opening remarks \cite{OPEN} I had shown a picture as an allegory on the future of high energy 
physics.  I am showing Fig.\ref{FUTURE2} again for a related reason. You notice the sun between 
the two rocks. Just looking at the picture without taking recourse to additional information you might have about these structures (in particular if you are Italian) you cannot tell if the sun is rising or setting. 
To me one of the most impressive aspects of experimental high energy physics is that when you give  
experimentalists resources and time it is most amazing what they can achieve. As an example just have a look at the `Blue Book' that was written during the planning and constructing of LEP with a lot of 
quality time spent on it by theorists as well and compare it with what was actually done -- which was so much more. This 
speaks most highly of the intellectual vigour of the field. 

Likewise I firmly believe that if a Super-B factory is built, a proper time span is provided to utilize it 
and young people can fully participate in shaping its program, we will learn even more than what we envision now. In that sense I am confident the picture shows a rising rather than a 
setting sun. 

\vspace{1cm}

{\bf Acknowledgments}

\noindent It always is a most gratifying experience to come to the Rome area for discussing  
and marveling about 
nature's puzzles with colleagues.  This was also true this time, and I am thankful for the organizers 
of this workshop for creating this opportunity. This work was supported by the NSF under grant PHY03-55098.

\end{document}